Enhancement in the Figure of Merit of p-type $Bi_{100-x}Sb_x$ alloys through multiple valence-band doping

Hyungyu Jin[1], Christopher M. Jaworski[1], and Joseph P. Heremans[1,2]

1. Department of Mechanical and Aerospace Engineering, The Ohio State University, Columbus, OH, USA
2. Department of Physics, The Ohio State University, Columbus, OH, USA

**ABSTRACT**

N-type $Bi_{100-x}Sb_x$ alloys have the highest thermoelectric figure of merit ($zT$) of all materials below 250K; here we investigate how filling multiple valence band pockets at the T and $\Box$-points of the Brillouin zone produces high $zT$'s in p-type Sn-doped material. This approach, theoretically predicted to potentially give $zT>1$ in Bi, was used in PbTe. We report thermopower, electrical and thermal conductivity (2 to 400K) measurements of single crystals with $12 \leq x \leq 37$ and polycrystals (x=50 - 90), higher Sb concentrations than previous studies. We obtain a 60% improvement in $zT$ to 0.13.

N-type $Bi_{100-x}Sb_x$ alloys are the materials with the highest thermoelectric figure of merit ($zT \sim 0.5$ at 100K[1]) below 200K; $zT=S^2\sigma T/\kappa$, where $S$ is the Seebeck coefficient or thermopower, $\sigma(=1/\rho)$ the electrical conductivity (resistivity) and $\kappa$ the thermal conductivity of a material. The performance of p-type material is not equivalent: the highest $zT$ of p-type $Bi_{100-x}Sb_x$ alloys reported thus far is about 0.08 at 200K for the composition of $Bi_{88.5}Sb_{7.5}Sn_4$.[2] Investigations of p-type $Bi_{100-x}Sb_x$ alloys have been made for the semiconducting compositions which correspond to 7-20% Sb,[2,3,4,5] the range of compositions that yields high $zT$ values in n-type material, but doped with various concentrations of the acceptor atom Sn. Because the nature of the valence bands is quite different from that of the conduction band, we extend here the study of the p-type material to the semi-metallic Sb-rich region, where p-type material has not been studied before, reaching x=37% Sn in single crystals and 50-90% in polycrystals; we report a 60% improvement of the maximum $zT$.

Contrasting with the experimental observations, Thonhauser et al.[6] predicted that $zT=1.4$ at room temperature can theoretically be reached if Bi was heavily doped p-type so that its Fermi level would fall about 0.25eV below the upper valence band (VB) edge (3.1 atomic % of Sn, if each Sn atom substituted for a Bi atom was to capture one electron). This shifts the Fermi level into the heavy VBs near the H-points (H stands for the Greek letter eta) inside the Brillouin zone, so that the hole Fermi surface consists of the T-hole pockets and saddle points near the heavy H-hole bands. This approach was shown to be successful in PbTe[7] using the hole levels near its Σ-point. In elemental Bi, the approach runs into a difficulty: at temperatures above 220K, Sn loses its activity as an acceptor,[8] due to a change in the relative band positions and Sn-impurity level with temperature.[9] We address that difficulty here by observing that the H-band moves up in energy with x in $Bi_{100-x}Sb_x$ alloys at $T = 0K$ (see Fig. 1, compiled from [10,11,12] and [13]; we could not compile a similar figure for higher temperatures, because not all the parameters are known above cryogenic temperatures[9]). Conceptually, the idea is as follows: the motion of the Bi bands with increasing $T$ is mostly due to thermal expansion effects. Because the lattice constant of $Bi_{100-x}Sb_x$ alloys decreases with increasing x, the latter effect can be viewed to the first order as compensating the effect of $T$. This suggests that one may dope the H-bands more effectively at high Sb concentration. Fig. 1 shows that at about 15-17% Sb, the H-band crosses over the light VB at the L-point and keeps increasing in energy while the T-band decreases to be located at lower energy than that of the H-band. Hence, for x>15-17% Sb, the VB nearest to the L-point conduction band becomes the H-band, and we can expect to dope it by introducing an acceptor impurity. We additionally dope the T- and L- bands, fulfilling the condition of Ref [6]. The concept is reinforced by the observation[12] that Sn remains as an acceptor at 300K in pure Sb. Sn is chosen over Pb based on the work of Noguchi et al.[14] We show experimentally that the concept works, although not well enough to reach $zT$ values near the theoretical prediction.

$Bi_{100-x}Sb_x$ single crystals with $12 \leq x \leq 37$ doped with 0.75 at % Sn (corresponding to roughly $\sim 5 \times 10^{19}$ atoms of Sn per $cm^3$), and polycrystalline samples with x = 50, 70, 80, and 90 at % were prepared for thermoelectric properties measurement. The work is organized as follows: first, we compare the transport properties between along the trigonal axis and in the trigonal plane, and show that the best performance occurs when the fluxes are in-plane. Then, we report the complete set of the $S$, $\rho$, $\kappa$, and the corresponding $zT$ of the single crystal samples. Finally, the relation between carrier concentration and thermopower, known as Pisarenko's relation, for all samples is constructed and interpreted based on the Sb concentration dependent band structure of $Bi_{100-x}Sb_x$ alloys.

Seven $Bi_{100-x-0.75}Sb_xSn_{0.75}$ single crystals were grown by a modified Bridgeman technique. Elemental Bi (5N), Sb (6N), and Sn (5N) were loaded into the quartz ampoules following the stoichiometric ratios and the ampoules were sealed at less than $10^{-6}$ torr. Then the samples were melted at 632°C in a tube furnace and slowly pulled out of the furnace at a rate of 0.13mm/hr. Because of the large segregation coefficient in the Bi-Sb phase diagram, it is known to be difficult to grow a homogeneous $Bi_{100-x}Sb_x$ single crystal, and the samples were annealed for 6 months at 255°C. In addition, four polycrystalline $Bi_{100-x-0.75}Sb_xSn_{0.75}$ samples with x=50, 70, 80, and 90 were prepared by ball-milling and cold pressing followed by 2 weeks sinter at 255°C. Powder XRD confirmed single crystal peaks and polycrystalline peaks as well as shift of the peaks as Sb concentration increases. The actual Sb concentration reported here for the samples was identified by X-ray diffraction using Vegard's law. Then the samples were cut into approximately 2.5 x 1.5 x 7mm$^3$ parallelepipeds for thermoelectric measurement. For the single crystal samples, the parallelepipeds were obtained in two different crystallographic directions: the long axis parallel to the trigonal axis and to the trigonal plane, respectively. *ρ(T)*, *κ(T)*, and *S(T)* were measured simultaneously in a quasi-static heater-and-sink configuration while slowly sweeping the temperature from 400K to 2K using the Thermal Transport Option (TTO) in a Physical Properties Measurement System (PPMS) by Quantum Design. Hall coefficient of each sample was measured using AC Transport Option in the PPMS at several different temperatures with the magnetic field up to 7T and oriented along the trigonal axis direction of the sample. Because the crystal symmetry of Bi and $Bi_{100-x}Sb_x$ alloys is such[15] that *ρ*, *κ* and *S* are isotropic in the trigonal plane even in the presence of a magnetic field aligned along the trigonal axis, we did not identify the specific in-plane binary or bisectrix crystallographic direction when we made measurements in that plane, and no optimization of *zT* is possible by further orienting the current and heat flux in the plane.

Figure 2 shows *κ(T)*, *S(T)*, *ρ(T)*, and *zT* of the $Bi_{81.05}Sb_{18.2}Sn_{0.75}$ sample in the trigonal plane and in the trigonal axis direction, respectively. The best *zT* in n-type $Bi_{100-x}Sb_x$ alloys is obtained along the trigonal axis direction[1] due to much higher mobility of electrons than holes in this direction. We show in Fig. 2d that for p-type material, in contrast, *zT* is higher in the trigonal plane. In-plane *ρ* and *S* are better than their equivalents along the trigonal axis. In spite of the less favorable *κ*, the resulting *zT* at 150K is about five times larger in the trigonal plane than perpendicularly to it. This effect is also due to the difference in mobility between electrons and holes, which here contributes to *zT* by two mechanisms. At 4K, the hole mobility in the trigonal plane is ~6 times higher than that along trigonal axis in single crystal bismuth.[16] This is reflected firstly in the fact that the electrical conductivity (*σ*) of p-type material in the trigonal plane will show ~6 times larger than along the trigonal axis, as observed experimentally in *ρ* in Fig. 2c at the lowest temperatures. But the anisotropy of the hole mobility also affects the thermopower, which is less compensated by the presence of minority electrons in the plane than along the trigonal axis. The total $S_i$ (i = 1 or 3 for the trigonal plane or the trigonal axis direction, respectively) is expressed as:

$$S_i = \frac{S^e \sigma_i^e + S^h \sigma_i^h}{\sigma_i^e + \sigma_i^h} \qquad (1)$$

where $\sigma_i^e$ and $\sigma_i^h$ are the partial contribution of electrons and holes to the total electrical conductivity in the corresponding direction, and $S^e$ and $S^h$ are the partial thermopower of electrons and holes, respectively. Because the mobility ratio of holes to electrons is the largest in the trigonal plane, the contribution of the $S^e\sigma_i^e$ term which is negative is minimized, and consequently, we can obtain the largest p-type thermopower in the plane (Fig. 2b). Therefore, due to the superior thermoelectric properties in the trigonal plane, we present the transport results of the samples in this direction only.

Thermoelectric properties in the trigonal plane of single crystal $Bi_{100-x}Sb_xSn_{0.75}$ samples with $11 \leq x \leq 37$ are presented in Fig. 3. – we do not report the temperature-dependences of the polycrystalline samples for lack of space, but will use the data at 77K later. $\kappa(T)$ (Fig. 3a) of all Sn-doped samples follow the classical behavior of undoped $Bi_{100-x}Sb_x$ alloys, but they have lower values especially at low temperatures. Below 20K, heat transport in $Bi_{100-x}Sb_x$ alloys is dominated only by phonons, and has a maximum near 10K. Fig. 3a shows that the amplitude of the peak in $\kappa(T)$ decreases with increasing Sb concentration, consistently with the effects of alloy scattering on the lattice thermal conductivity ($\kappa_L$). All samples exhibit similar temperature dependence in $\rho(T)$ (Fig. 3b) that is typical in heavily doped small band-gap semiconductors. Increasing Sb concentration generally decreases $\rho$, consistently with the increasing band overlap with increasing x (Fig. 1).[17] One exception is the sample with 22.9% Sb which has the lowest $\rho$ among the samples despite of its intermediate Sb concentration, a fact that is noted without explanation. For the samples with relatively low Sb concentration (x≤19.5), there is a point where $S$ changes its sign from p-type to n-type below 400K (Fig. 3c). This turnover point tends to move to higher temperature as the Sb concentration increases, as expected from the arguments about the Sb compensating for the temperature-dependence of the band structure made in the introduction. Samples with higher Sb concentration (x>19.5) show positive thermopower over the measured temperatures. However, the magnitude of the *S* decreases rapidly as Sb concentration exceeds 30%, indicating a steady increase in carrier concentration with increasing Sb content, which is consistent with the increase in overlap between H and L-point bands shown in Fig. 1 for x>30, and the concomitant increase in hole concentration. The largest p-type thermopower is about +60 µV/K obtained for 11.6% Sb sample at 150K. Lastly, the inset in Fig. 3c shows that the best *zT* is obtained for the sample with 22.9% Sb at 240K reaching *zT*=0.13. While this value is admittedly low, it is the best p-type *zT* reported in this system. According to Fig. 1, the H band crosses over the conduction band at the L-point at about 22% Sb, and therefore the alloy turns into semimetal again so that this concentration could be one of the most favorable points for doping the H-bands as well as the additional T- or L-bands. The transport results for $Bi_{76.35}Sb_{22.9}Sn_{0.75}$ sample seem to confirm the singularity of this composition.

Fig. 4a shows a plot of thermopower as a function of hole concentration (Pisarenko's plot) obtained from the Hall measurements at T=80K for all of the samples in this study. The solid line and the dashed line are calculated[8] for the VBs centered at the T-point and at the H-points of the Brillouin zone, respectively, assuming that the density of states in those bands is not a function of x, and also assuming acoustic phonon scattering. There are three distinctive sample groups, Figs. 4b, 4c and 4d, in the plot classified by the relative position of the different VB pockets vis-à-vis the Fermi level ($E_F$, solid line) calculated for the H-bands. Readers are also advised to refer to Fig. 1 in order to follow how the band structure evolves as Sb concentration increases. In the first group (Fig. 4b), 3 set of pockets (at L, T and H-points) of the VB have their extrema located within 25 meV of each-other. Considering that 0.75% Sn corresponds to heavy doping in Bi,[18] there is a possibility that 2 or 3 VBs at different symmetry points in Brillouin zone are doped at the same time even though their extrema are located at different energies: all their extrema are above $E_F$.. The thermopower $S$ is then given by Eq. (1), and is increased over the partial $S$ of the H-bands, since all doped bands have partial thermopowers of the same polarity. With further increase in Sb concentration, the T-valence band edge goes deeper in energy and moves vis-à-vis $E_F$, so that only the H-bands are doped (Fig. 4c). This picture corresponds to the second group of 30%, 37%, and 50% Sb samples, whose thermopower falls on the solid line in Fig. 4a. Lastly, in the three samples with the highest Sb concentrations (70%, 80%, and 90%), the energy overlap between the H-bands and the conduction bands at the L-points is large enough for $E_F$ to cross the bottom of the L-conduction bands (Fig. 4d) and the samples become more compensated. The presence of minority electrons at 80K, in spite of heavy p-type doping with 0.75% Sn, in turn decreases $S$ below that of the H-bands. We conclude that p-type doping is most effective when it involves more than one type of VB (the case of Fig. 4b), and it is in this category that the best $zT$'s in this study are found (inset in Fig. 3c).

To summarize this work, we explore doping of $Bi_{100-x}Sb_x$ alloys to enhance the p-type $zT$ by populating multiple valence bands including the heavy bands near the H-points. P-type alloys have their highest $zT$ in the trigonal plane, unlike n-type alloys, due to the anisotropy of the hole mobility. Samples with x>19.5at% Sb remain p-type to 300K, unlike those with less Sb, because the temperature-dependence of the relative positions of the impurity atom level and the bands is partially compensated by the x-dependence of the bands. One sample with 22.9% Sb has $zT$ = 0.13 at 240K which is the highest so far in this system. Based on the Pisarenko's plot and the band structure of $Bi_{100-x}Sb_x$ alloys, we conclude that the *zT* of p-type BiSbSn system is indeed improved by doping multiple valence bands, consistently with the theory of Thonhauser et al.,[6] although *zT* of p-type BiSb is significantly lower than that calculated, or than that of n-type alloys.

Acknowledgement: this work is supported by the Air Force Office of Scientific Research MURI "Cryogenic Peltier Cooling", contract FA9550-10-1-0533.

**FIGURE CAPTIONS**

FIG. 1.

(Color online) Dependence of the energies of the various band extrema on composition x in $Bi_{100-x}Sb_x$ alloys at $T=0K$ (compiled from Refs. 9-12). At the L-points of the Brillouin zone the symmetric and antisymmetric bands are inverted between elemental Bi and Sb, leading to a Dirac point near x=5 at % fraction. The T-point valence band is the upper valence band of the semimetal Bi, but in elemental Sb, the holes are inside the Brillouin zone in three distorted ellipsoidal pockets near the H-points. This band becomes the upper valence band for x>18. As in most solids, temperature can make band extrema shift by 100 meV between liquid nitrogen and room temperatures, an effect that is secondary in wide-gap semiconductors, but has a very important relative influence on the present diagram.

FIG. 2.

(Color online) Comparison of thermoelectric properties of single crystal $Bi_{81.05}Sb_{18.2}Sn_{0.75}$ sample between in the trigonal plane and along the trigonal axis directions. (a) thermal conductivity, (b) Thermopower $S$, (c) electrical resistivity $\rho$, (d) Figure of merit $zT$ as functions of temperature. Points are experimental data, lines are added to guide the eye.

FIG. 3.

(Color online) Thermoelectric properties of single crystal $Bi_{100-x}Sb_xSn_{0.75}$ samples for x = 11.6, 18.2, 19.5, 22.9, 26.5, 30, and 37 from 2K to 400K. All properties were measured with the heat flux and current in the trigonal plane. (a) thermal conductivity $\kappa$, (b) electrical resistivity $\rho$, (c) Thermopower $S$ as functions of temperature. The insert shows the figure of merit $zT$ as a function of temperature. Points are experimental data, lines are added to guide the eye.

FIG. 4.

(Color online) (a) Pisarenko's plot of thermopower versus hole concentration at $T$=80K for all measured samples in this study. Solid line and dashed line are calculated for the valence bands at the H-points and the valence band at the T-point, respectively. Points are experimental data. Carrier density of each sample was obtained from Hall measurement. (b), (c), (d) Schematic band diagrams for the three different Sb concentration regimes shown. Relative position of each band and the position of Fermi energies ($E_F$) are not to scale.

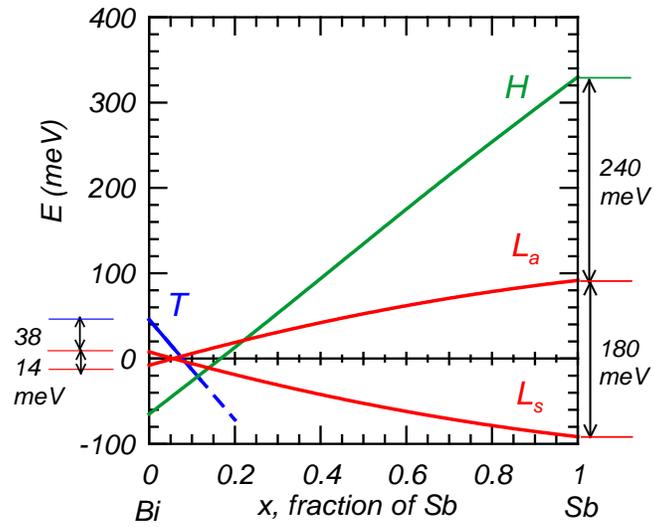

FIG. 1.

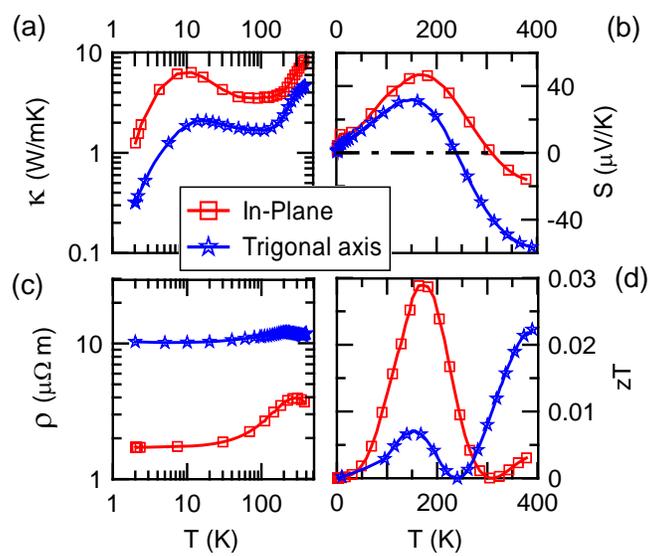

FIG. 2.

H. Jin et al.

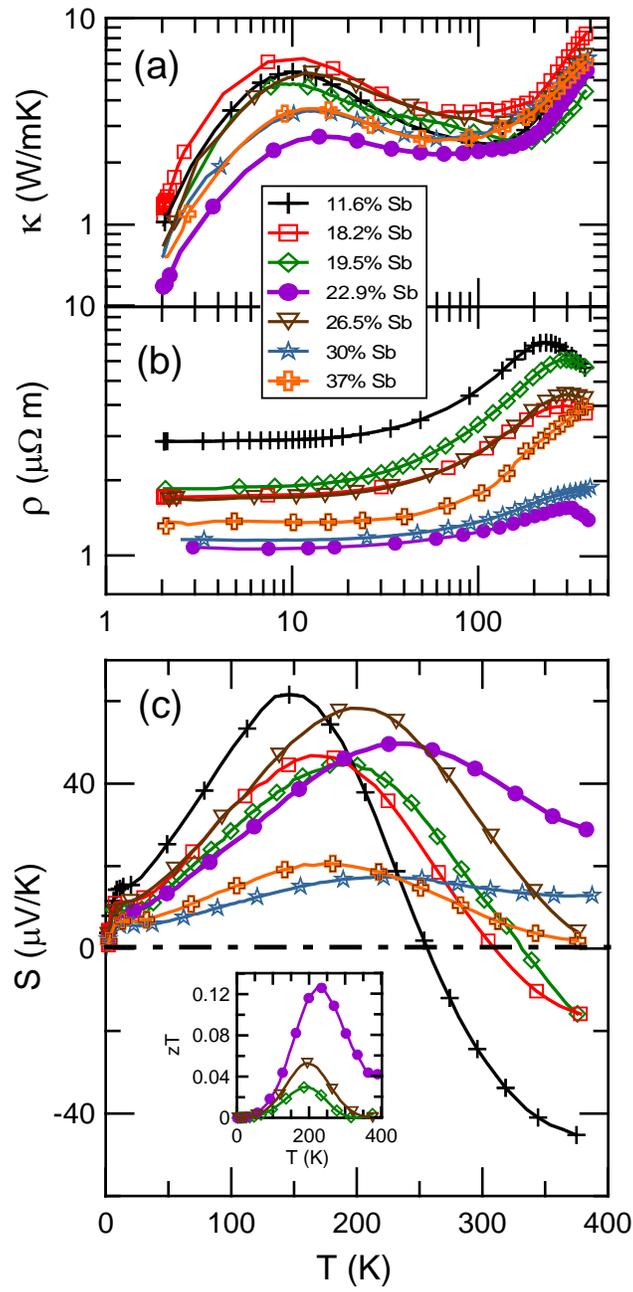

FIG. 3.

*H. Jin et al.*

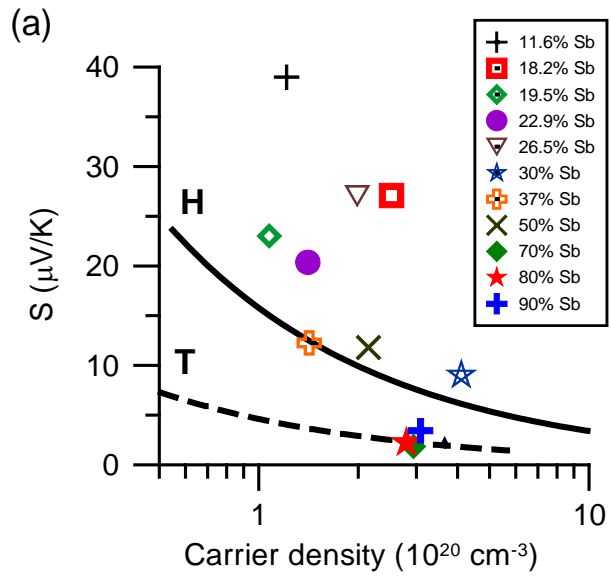

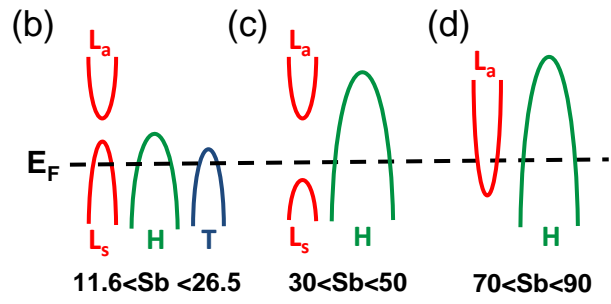

FIG. 4.

*H. Jin et al.*

# REFERENCES

[1] G. E. Smith and R. Wolfe, J. Appl. Phys. **33**, 841 (1962).

[2] Y. S. Hor and R. J. Cava, J. Alloys and Compounds **479**, 368 (2009).

[3] W. M. Yim and A. Amith, Solid-State Electronics **15**, 1141 (1972).

[4] M. Sakurai, T. Ono, I. Yoshida and S. Tanuma, 17th Int. Conf. on Thermoelectrics, 134 (1998).

[5] J. W. Sharp, G. S. Nolas and E. H. Volckmann, Mat. Res. Soc. Symp. Proc. **478**, 91 (1997).

[6] T. Thonhauser, T. J. Scheidemantel and J. O. Sofo, Appl. Phys. Lett. **85**, 588 (2004).

[7] Y. Pei, X. Shi, A. LaLonde, H. Wang, L. Chen and G. J. Snyder, Nature **473**, 66 (2011).

[8] J. P. Heremans and O. P. Hansen, J. Phys. C: Solid State Phys. **16**, 4623 (1983).

[9] J. P. Heremans and O. P. Hansen, J. Phys. C: Solid State Phys. **12**, 3483 (1979).

[10] N. B. Brandt, R. Hermann, G. I. Golysheva, L. I. Devyatkova, D. Kusnik, W. Kraak and Ya. G. Ponomarev, Sov. Phys. JETP **56,** 1247 (1982).

[11] N. B. Brandt, E. A. Svistova and M. V. Semenov, Sov. Phys. JETP **32,** 238 (1971).

[12] S. Epstein and H. J. Juretschke, Phys. Rev. **129**, 1148 (1963).

[13] B. Lenoir, M. Cassart, J.-P. Michenaud, H. Scherrer and S. Scherrer, J. Phys. Chem. Solids **57**, 89 (1996).

[14] H. Noguchi, H. Kitagawa, T. Kiyabu, K. Hasezaki and Y. Noda, J. Phys. Chem. Solids **68**, 91 (2007).

[15] Y. C. Akgöz and G. A. Saunders, J. Phys. C. Solid State Phys. **8**, 1387 (1975) and **8**, 2962 (1975).

[16] R. Hartman, Phys. Rev. **181**, 1070 (1969).

[17] J. P. Issi, Aust. J. Phys. **32**, 585 (1979).

[18] C. Uher, J. Phys. F: Metal Phys. **9**, 2399 (1979).